\documentstyle{aipproc}
\def\laq{\ \raise 0.4ex\hbox{$<$}\kern -0.8em\lower 0.62
ex\hbox{$\sim$}\ }
\def\gaq{\ \raise 0.4ex\hbox{$>$}\kern -0.7em\lower 0.62
ex\hbox{$\sim$}\ }
\def\half{\hbox{\magstep{-1}$\frac{1}{2}$}}

\def\NPB{{\em Nucl. Phys.} B}
\def\PLB{{\em Phys. Lett.}  B}
\def\PRL{{\em Phys. Rev. Lett. }}
\def\PRD{{\em Phys. Rev.} D}

\begin{document}

\title{Dark Matter in\\ Models of String Cosmology}

\author{Ram Brustein $^1$ and Merav Hadad $^2$} 
\address{$^1$ Department of Physics,
Ben-Gurion University,
Beer-Sheva 84105, Israel\\
 $^2$School of Physics and Astronomy,
Tel Aviv University, Tel Aviv 69978, Israel\\
email: ramyb@bgumail.bgu.ac.il, meravv@post.tau.ac.il}

\maketitle
\begin{abstract}

The origin of dark matter in the universe may be 
weakly interacting scalar particles produced by amplification of quantum
fluctuations during a period of dilaton-driven inflation. We present two
interesting cases,  the case of small fluctuations, and the resulting
nonthermal spectrum, and  the case of large fluctuations of  a field
with a periodic potential, the QCD axion.

 \end{abstract}

We consider particle production in models of string cosmology which
realize the pre-big-bang (PBB) scenario \cite{sfd}. In this scenario the
evolution of the universe starts from a state of very small curvature and
coupling and then undergoes a long phase of dilaton-driven inflation (DDI)  and at some
later time joins smoothly standard  radiation dominated (RD) 
cosmological evolution,
thus giving rise to a singularity free inflationary cosmology.
Particles are produced during DDI phase by the standard mechanism of amplification of
quantum fluctuations \cite{mukh}. Some debate about the naturalness of initial
conditions and whether PBB models actually solve the  cosmological
problems has been taking place \cite{ic}. The smooth transition from DDI
to RD is also not completely solved, for ideas about how this may come about
see \cite{exit}. We will take a more phenomenological approach, 
and concentrate on interesting consequences of the models which 
do not depend on the detailed resolution of these issues, 
 assuming that a resolution exists.

In the simplified model of background evolution we adopt, the evolution of the
universe is divided into four distinct  phases  with specific (conformal)
time dependence of the scale factor of the universe $a(\eta)$ and the dilaton
$\phi(\eta)$.  We assume throughout an isotropic
and homogeneous four dimensional flat universe, described by a 
FRW  metric. All
other scalar fields  are assumed to have a trivial vacuum expectation value 
during the inflationary phase.

We have computed spectrums of produced particles for the models 
described above \cite{bh1}. We have solved a linear 
perturbation equation, 
$
\chi_k''+\left(  k^2+M^2a^2-s''/s \right)\!\chi _k=0,
$
where 
$
s(\eta )\equiv a(\eta )^m e^{l\phi (\eta )/2}=a_s^me^{l\phi _s/2}\left( 
\eta/\eta _s\right) ^{1/2-n_s},
$
imposing initial conditions corresponding to normalized vacuum fluctuations.
The parameter $m$ depends on the spin of the particle and $l$ depends on
its coupling to the dilaton.
Similar calculations have been performed by several groups  
and the results agree \cite{cllw}, 
whenever a comparison was possible.

We consider weakly interacting scalar particles, 
abundant in string
theory and supergravity. For scalar fields $m=1$, and we will consider 
for concreteness the  values,
$l=-1,0,1$ corresponding, respectively, to moduli,
Ramond-Ramond axions, and Neveu-Schwartz axions. We will assume that 
the produced particles interact so weakly, that their interactions
and decay are not sufficient to alter the primordial spectrum substantially.
The particles we have in mind have typically gravitational strength 
interactions, which is definitely weak enough to satisfy our assumption, and
 their masses are below a fraction of an eV.

A typical spectrum of a light scalar may be divided, at a given time, 
into three physical momentum (PM) regions:
$i)$ The massless region, $\omega_S>\omega>M$, 
in this region particles are relativistic. 
$ii)$ The ``false'' massive region,  $M>\omega>\omega_m$, where
$\omega_m=\omega_1(M/M_s)^{1/2}$. 
In this region particles are NR, 
but have reentered the horizon as relativistic modes.
$iii)$ The ``real'' massive region $\omega_m>\omega$. 
In this region particles are non-relativistic (NR), 
and have reentered the horizon as NR modes. Note that 
PM redshift as the universe expands, and therefore 
boundaries of  regions change in time,
\begin{equation}
\! \frac{d\Omega }{d\ln \omega}\! = \!
{\cal N}
g_1^2\!\left(\frac{g_1}{g_S}\right)^{2l}\!\!\cases{
\left( \frac {\omega}{\omega_S}\right) ^x  
 &$\omega_S\!>\!\omega\!>\!M$    \cr 
\frac {M}{\omega_S}\left( \frac
{\omega}{\omega_S}\right) ^{x-1}  
 & $M\!>\!\omega\!\!>\omega_m$    \cr 
\!\!\frac {M}{\omega_1}\sqrt{\frac {M}{M_s}}\! 
\left( \frac {\omega}{\omega_S}\right) ^x  
  &$\omega_m\! >\! \omega$,   }
\label{chspectrum}
\end{equation}
where 
$
x\equiv 2+2\alpha +l\beta 
$, and ${\cal N}$ is a numerical factor, estimated in \cite{bh1}, which
we will set to unity in what follows. Parameters appearing in 
eq.(\ref{chspectrum}) are,
the string scale $M_s$, $z_S$ the total redshift during the string phase,
$g_S$ and $g_1$, the string coupling at the beginning and end 
of the string phase, and  $\omega_1$, the PM (today), corresponding
to the end of the string phase, estimated in \cite{peak} to be $\omega_1\sim
10^{-5}eV$, and the PM $\omega_S= \omega_1/z_S$, the PM (today)
corresponding to the end of the DDI phase. 
In  (\ref{chspectrum})  we have assumed no substantial 
late entropy production has occurred. 
In general \cite{peak}, 
the effect of late entropy production is to further 
redshift the physical frequencies as $\sim (1-\delta s)^{1/3}$
 where $\delta s$ is the fraction of produced entropy,
 and, more importantly, to dilute the contribution of 
 modes which are already inside the horizon
 by a factor $(1-\delta s)^{4/3}$. 
If a substantial amount of entropy is produced
below $T\sim M_s/\sqrt{z_S}$, 
then spectrum (\ref{chspectrum}) is no longer a good 
approximate spectrum.

The first example we look at is an example of small field fluctuations.
The produced spectrum in this case is nonthermal, and may lead to 
an interesting case of cold and hot dark matter from the same source 
\cite{bh3}.
We look at generalized axions ($l=1$) with masses 
below $.1$ eV in a cosmological model
described in \cite{sfd}. In this model, $d=3$ spatial dimensions are
expanding and $n=6$ spatial dimensions are contracting, leading to
 $\alpha =-2/\left( d+n+3\right) =-1/6$ and $\beta
=-4d/\left( d+n+3\right) =-1$.
For this specific model $x=2/3$, 
$ \Omega_{REL}\simeq g_1^2\left( \frac{g_1}{g_S}\right)^2$, 
$\Omega _{NR}\simeq g_1^2\left( \frac{g_1}{g_S}\right)^{2}
\frac {M}{\omega_S}\left( \frac{\omega_m}{\omega_S}\right) ^{-1/3}$
Taking $10^{-10}\hbox{\rm eV}<M<10^{-2}\hbox{\rm eV},$ for which the above
 condition is comfortably satisfied, we observe that 
the ratio $\Omega _{REL}:$ $\Omega_{NR}$ at the start 
of structure formation era can vary in a range from
 well above unity  to well below unity, corresponding to hot, mixed and
 cold dark matter. 
For example, choosing $g_1=.1$ and $g_S=.01$ 
if the axion's mass is $10^{-10}\hbox{\rm eV}$, and for 
$z_S\sim 2\times 10^{4}$ we
get $\Omega _{REL}:$ $\Omega _{NR}=$ $1:1$ with both energy densities
being near critical, and if we choose
$g_1=.1$ and $g_S=.03$, making $\Omega_{REL}\simeq .1$,
and if $z_S\sim 10^{6}$ we 
obtain
$\Omega _{REL}:$ $\Omega _{NR}=1:10$, with $\Omega _{NR}\simeq 1$. 
Note that  ratio   depends on $z_S^{-2/3} M^{-5/6}$, 
so the previous examples correspond to a range of allowed values.

We now turn to the case of large field fluctuations. The spectrum
in this case can be very different from that given by the naive result
eq.(\ref{chspectrum}). We look at the model independent axion 
\cite{witten},
$l=1$, assuming that it is the QCD axion \cite{pq} in a model of background evolution
in which d=3 spatial dimensions expand, and $n=6$ spatial dimensions are
fixed, leading to  $x=3-2\sqrt{3}\sim-0.46$ \cite{bh2}.
 
Because of the negative exponent the spectrum is dominated by the lowest 
PM entering the horizon at a given time. The most interesting situation
is when the axion potential turns on when
the universe  cools down to QCD temperatures. If we try to approximate the axion
potential by a quadratic potential, leading to the result 
(\ref{chspectrum}), we
encounter a puzzle. The axion energy density becomes formally divergent as soon
as the axion potential turns on! 
Once the potential is generated, all the low frequencies
reenter the horizon at once, so to obtain the total 
energy density inside the horizon we need to integrate it from the
minimal amplified frequency $\omega_{min}$, which is either zero, if the
duration of the dilaton-driven phase is infinite, or exponentially small if the
duration is finite but large. The lower frequency part of the
spectrum yields a divergent contribution, proportional to 
${\omega_{min}}^{3-2\sqrt{3}}$. This
result does not make sense.

The resolution of the puzzle depends crucially on the periodic nature of
axion potential $V(\psi)=\half V_0\left(1-\cos(\frac{\psi}{\psi_0})\right)$. 
This point was first understood by Kofman and Linde \cite{koflinde}, 
and we have adopted their ideas to our particular situation. 
First, the total  potential energy is limited to $V_0$ and does not continue to increase
indefinitely as the axion field increases, providing a ``topological cutoff" on
the total axionic energy density and as important, large fluctuations in the
axion field are also  ``topologically cutoff", producing exponentially small
energy density perturbations.  Large fluctuations lead to a uniform
distribution of the axion field inside the horizon, with very small statistical
fluctuations.

Using  completely standard arguments \cite{cosmaxion}, we may obtain a
bound on the mass of the axion
$m_a$  (or equivalently on $\psi_0$) by requiring that the energy
density  in the coherent axion oscillations  be subcritical at the beginning of
matter domination epoch. This requirement leads to the  
standard bounds on the axion mass,
$
\Omega_a h^2 \sim \frac{10^{-6}eV}{m_a},
$
where  $h$ is todays Hubble parameter in units of 100 km/Mpc/sec.
Requiring subcritical $\Omega_a$ leads to the standard bound on $\psi_0$,
$\psi_0\laq 10^{12} GeV$ and $m_a\gaq10^{-6} eV$.
In string theory, natural values of $\psi_0$ are approximately $10^{16}
GeV$, which, if taken at face value, would lead to overclosure of the universe
with axions many times over. Two possible resolutions have been suggested 
\cite{banksdine} 
to allow our universe to reach its old age of today.

If dark matter in the universe is indeed made of light particles
with gravitational strength interactions its detection in current 
direct searches is extremely difficult, and will probably require new methods
and ideas. 
 
{\bf Acknowledgments }

Work  supported in part by the  Israel
Science Foundation.


\begin{references}

\bibitem{sfd}
G. Veneziano, \PLB 265 (1991) 287;
 M. Gasperini and G. Veneziano, 
{\em Astropart. Phys.} 1 (1993)  317.

\bibitem{mukh} 
N. Birrell and P. Davies, {\em Quantum fields in curved space}, 
Cambridge University Press, 1984;
V. Mukhanov, A. Feldman and R. Brandenberger,
{\em Phys. Rep. } 215 (1992) 203.

\bibitem{ic}
G. Veneziano, \PLB 406 (1997) 297; 
E. Weinberg and M. Turner, \PRD 56 (1997) 4604; 
M. Maggiore and R. Sturani, \PLB 415 (1997) 335; 
A. Buonanno, et al, \PRD 57 (1998) 2543;  
A. Buonanno, T. Damour and G. Veneziano, hep-th/9806230.
 

\bibitem{exit}  
R. Brustein and G. Veneziano, \PLB 329 (1994) 429; 
N. Kaloper, R. Madden and K. Olive, \NPB 452 (1995) 677; 
M. Gasperini, M. Maggiore and G. Veneziano, \NPB 494 (1997) 315;
R. Brustein and R. Madden, \PLB 410 (1997) 110; \PRD 57 (1998) 712;
hep-th/9901044; G. veneziano, hep-th/9902126; S.~Foffa, M.~Maggiore and 
R.~Sturani,
hep-th/9903008.

\bibitem{bh1}
R. Brustein and M. Hadad, \PRD 57 (1998) 725.

\bibitem{cllw}
E. Copeland, et al,    
\PRD 56 (1997) 874; E. Copeland, J. Lidsey, D. Wands, 
\NPB 506 (1997) 407;  hep-th/9809105;
A. Buonanno, et al, JHEP 01 (1998) 004;
M. Giovannini,  hep-th/9809185. 

\bibitem{peak}
R. Brustein, M. Gasperini and G. Veneziano, \PRD 55 (1997) 3882.


\bibitem{bh3}
R. Brustein and M. Hadad, hep-ph/9810526, to appear in \PRL.

\bibitem{witten} E. Witten, \PLB 149  (1984) 351;  \NPB 268 (1986) 79;
 K. Choi and J.  Kim, \PLB 154 (1985) 393;  \PLB 165 (1985) 71.


\bibitem{pq} R. Peccei and H.  Quinn, \PRL  38 (1977) 1440;
S. Weinberg, \PRL 40 (1978) 223; 
F. Wilczek, \PRL 40 (1978) 279.

\bibitem{bh2}
R. Brustein and M. Hadad, \PLB 442 (1998) 74.





         

\bibitem{koflinde}
A. Linde,\PLB 158  (1985) 375;       
L.  Kofman, \PLB 173 (1986) 400; 
L. Kofman and  A. Linde,  \NPB 282 (1987) 555.

          
\bibitem{cosmaxion} J. Preskill, M. Wise and F. Wilczek, \PLB 120 (1983) 127;  
L. Abbott and P. Sikivie, \PLB 120 (1983) 133;  
  M. Dine and W. Fischler, \PLB 120 (1983) 137.

\bibitem{banksdine} T. Banks and M. Dine, \NPB 479 (1996) 173;
 T. Banks and M. Dine, \NPB 505 (1997) 445;
K. Choi, \PRD 56 (1997) 6588.

\end{references}
\end{document}